%% file: main.tex
\documentclass[runningheads]{llncs}
\usepackage{graphicx}
\usepackage{subcaption}
\usepackage{amsmath}
\usepackage{hyperref}
\usepackage{cleveref}
\usepackage{amsfonts} 
\usepackage{amssymb} 
\usepackage{booktabs}
\usepackage{caption}
\usepackage{float}
\usepackage{rotating}

\begin{document}
\title{HEMIT: H\&E to Multiplex-immunohistochemistry Image Translation with Dual-Branch Pix2pix Generator}
\titlerunning{HEMIT}
%

\author{Chang Bian\inst{1} \and
Beth Phillips \inst{2} \and
Tim Cootes\inst{1} \and
Martin Fergie\inst{1}
}

\authorrunning{Bian et al.}
%
%
\institute{Division of Informatics, Imaging and Data Sciences, School of Health Sciences, University of Manchester\and
Division of Cancer Sciences, University of Manchester
\\
\email{chang.bian@postgrad.manchester.ac.uk}}

\maketitle              
\begin{abstract}
Computational analysis of multiplexed immunofluorescence histology data is emerging as an important method for understanding the tumour micro-environment in cancer. This work presents HEMIT, a dataset designed for translating Hematoxylin and Eosin (H\&E) sections to multiplex-immunohistochemistry (mIHC) images, simultaneously featuring DAPI, CD3, and panCK markers. Distinctively, HEMIT's mIHC images are multi-component and cellular-level aligned with H\&E, enriching supervised stain translation tasks. To our knowledge, HEMIT is the first publicly available cellular-level aligned dataset that enables H\&E to multi-target mIHC image translation. This dataset provides the computer vision community with a valuable resource to develop novel computational methods which have the potential to gain new insights from H\&E slide archives.

We also propose a new dual-branch generator architecture combining residual Convolutional Neural Networks (CNNs) and Swin Transformers, with a feature map fusion module to integrate information from both branches. This architecture achieves superior translation outcomes compared to other popular algorithms. Evaluations on the HEMIT dataset show it outperforms pix2pixHD, pix2pix, U-Net, and ResNet, with the highest scores in Structural Similarity Index Measure (SSIM), Pearson correlation (R), and Peak Signal-to-Noise Ratio (PSNR). Additionally, we discuss the limitations of commonly used metrics in certain stain translation scenarios and provide recommendations for future use.

Additionally, we also designed a downstream analysis to further validate the quality and utility of the generated mIHC images from a clinical-focused perspective. These results set a new benchmark in the field of stain translation tasks. 

The proposed dataset can be accessed at:\url{https://github.com/BianChang/HEMIT-DATASET}.

\end{abstract}
\input{sec/1_intro}

\input{sec/2_HEMITDataset}

\input{sec/3_ProposedMethod}

\input{sec/4_Experiments}
\input{sec/5_Conclusions}

%
%

%
%
%
\bibliographystyle{splncs04}
\bibliography{main}
\end{document}

%% file: sec/1_intro.tex
\section{Introduction}
\label{sec:intro}

Multiplex immunohistochemistry/immunofluorescence (mIHC/IF) technologies have revolutionized the study of the tumor microenvironment (TME), providing critical insights for cancer research and immunotherapy \cite{tan2020overview,bian2024integrating,wang2021novel,wang2022graph,an2021fast}. These technologies enable the detection of multiple markers in a single tissue slice, revealing intricate spatial interactions. However, the complexity and cost of mIHC limits its accessibility \cite{stack2014multiplexed}. Deep learning offers a promising solution to these challenges, with recent advancements in stain translation and virtual staining allowing for better exploitation of existing H\&E slides \cite{burlingame2018shift,jen2021silico,rivenson2019virtual,christiansen2018silico,weinstein2013cancer,bian2023transformer,bian2021computational,peng2022deep}.

Capitalizing on the ease and cost-efficiency of producing H\&E images, this paper aims to develop an image-to-image translation method that can convert H\&E images into their corresponding mIHC counterparts. 

Image-to-image translation algorithms are well-established across a range of image analysis domains. Specifically, pix2pix \cite{isola2017image} employs a conditional generative adversarial network (cGAN) approach for paired images, adeptly generating high-fidelity translations. Building upon this, pix2pixHD \cite{wang2018high} has demonstrated significant outcomes with high-resolution paired images. Notably, a recent study \cite{liu2022bci} introduced a multi-scale loss term, improving the translation performance for HER2 images. Furthermore, other methods \cite{liu2017unsupervised,kong2021breaking,menze2014multimodal} have been designed specifically for image-to-image translation tasks, underscoring the breadth and depth of research in this domain.

High-quality datasets are essential for effective supervised image translation, yet the realm of pathological image translation lacks comprehensive resources. The BCI dataset \cite{liu2022bci} is proposed for H\&E to HER2 IHC image translation. However, it is limited in use as staining is performed on consecutive tissue sections so there is no cell-to-cell mapping across stains. Further, it is limited to one target stain (HER2) unlike mIHC where multiple stains can be predicted simultaneously. In response, we present HEMIT: A dataset for H\&E to mIHC Translation. This is the first publicly-available cellular-level aligned dataset for stain translation. Concurrently, we propose a specialized method tailored for this task which we compare to various state-of-the-art image translation algorithms. This work makes the following contributions:
\begin{enumerate}
    \item  Introduction of HEMIT: a paired dataset for H\&E to mIHC image translation. To the best of our knowledge, HEMIT is the first publicly available cellular-level aligned dataset that enables H\&E to multi-target mIHC image translation.
    \item Development of a SwinTransformer-CNN-based dual-branch pix2pix strategy to convert H\&E images into mIHC versions. Our methodology assimilates both global information and spatial details, culminating in superior outcomes, setting benchmark results for the proposed dataset.
    \item A thorough empirical evaluation was conducted on the newly introduced HEMIT dataset, setting a benchmark for future investigations within the research community. We also discussed on the limitations of commonly used metrics in certain scenarios of stain translation. This discussion provides insights into the appropriateness and applicability of these metrics for different use cases.
    \item Further downstream analysis from clinical focused perspective, employing QuPath \cite{bankhead2017qupath}, substantiated the quality and utility of the images produced. This establishes a solid foundation for subsequent studies and applications, encouraging a deeper exploration into the dataset's potential. 
\end{enumerate}

%% file: sec/2_HEMITDataset.tex
\section{HEMIT Dataset}

\begin{figure}
  \centering
    \includegraphics[width=0.8\linewidth]{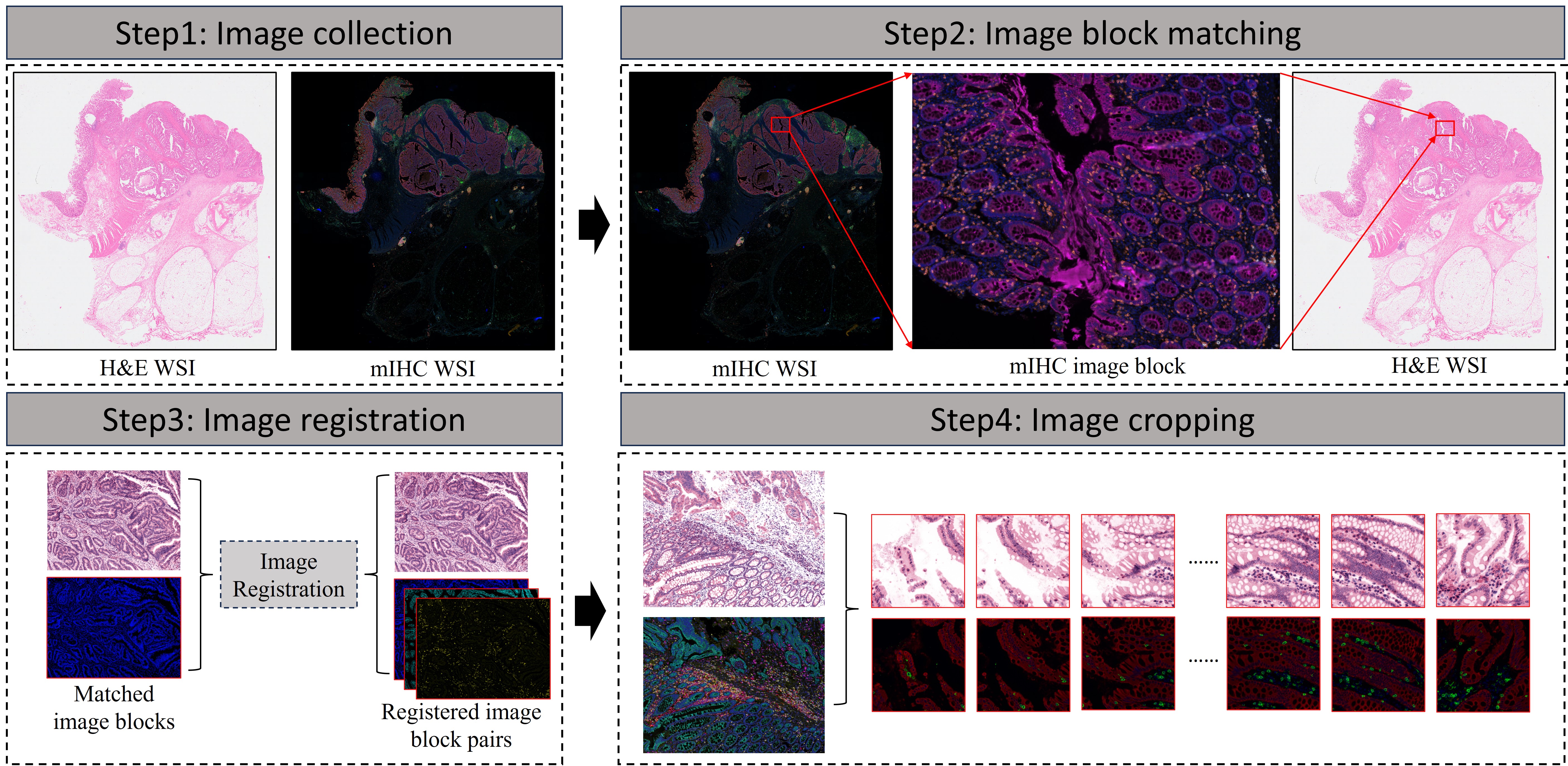}
  \caption{Overview of dataset processing pipeline of the HEMIT dataset.}
  \label{fig:dataset_overview}
\end{figure}

We present HEMIT: A cellular-level aligned dataset for H\&E to mIHC Image Translation. A schematic of our dataset's construction pipeline is provided in \cref{fig:dataset_overview}a. The proposed dataset can be accessed at: \url{https://github.com/BianChang/HEMIT-DATASET}.

\subsection{Data Collection}

HEMIT's raw data is sourced from ImmunoAIzer \cite{bian2021immunoaizer} which we have adapted to make it suitable for the computer vision community. Notably, HEMIT distinguishes itself from other datasets: the H\&E and mIHC slide pairs are derived from the identical tissue section, not consecutive slides. Consecutive slides often suffer from misalignment and variations in tissue morphology due to differences in the depths at which sectioning happens. This may affect the network's ability to learn accurate feature mappings between H\&E and mIHC stains. This misalignment leads to errors in translating the stains, which can degrade the model's performance and reliability in medical imaging applications. Examples of this misalignment are shown in \cref{fig:alignment}(b) using examples of existing datasets that rely on consecutive sections \cite{liu2022bci,li2023adaptive}. The tissue of HEMIT was first stained with mIHC protocols and then bleached before the H\&E staining. This feature leads to higher fidelity and alignment between the matched image pairs and better translation outcomes \cite{burlingame2020shift}. Example images demonstrating accurate pixel-level alignment of the HEMIT dataset and given in \cref{fig:alignment}(a).

In the mIHC images, three pivotal cell type identification markers are incorporated: DAPI to signify cell nuclei, pan-cytokeratin (panCK) to highlight tumor regions, and CD3 to pinpoint T cells—all of which are integral to TME analysis. These specific markers serve as the foundation for our H\&E to mIHC stain translation benchmark. We have collated a selection of publicly available datasets that leverage H\&E images for predicting IHC expressions. A comparative overview of their characteristics is presented in \cref{tab:datasets}. 

\begin{table}
\caption{Summary of publicly available datasets}\label{tab:datasets}
\centering
\begin{tabular}{|l|l|l|l|l|}
\hline
Datasets & Staining Types & Sectioning Approach & IHC/mIHC Markers & Ground Truths\\
\hline
HEMIT & H\&E \& mIHC & same slide & DAPI, CD3, panCK & cellular level\\
BCI \cite{liu2022bci} & H\&E \& IHC & consecutive slides & HER2 & structural level \\
HEROHE \cite{conde2022herohe} & H\&E & H\&E only & Clinical HER2 status & slide level\\
MIST \cite{li2023adaptive} &  H\&E \& IHC & consecutive slides & HER2, Ki67, PR, ER & structural level \\ 
\hline
\end{tabular}
\end{table}

\subsection{Data Preprocessing}

Despite both H\&E and mIHC staining being performed on the same tissue slide, the re-staining and scanning processes mean that the captured images do not align perfectly. We employ a 2-step registration process to ensure cellular-wise alignment of the image pairs which is crucial for optimal training performance. All registration steps are conducted under 40 times magnification. Implementation details of the registration process are given in the supplementary material.

\begin{figure}
  \centering
  \begin{tabular}{c}
    \includegraphics[width=1\linewidth]{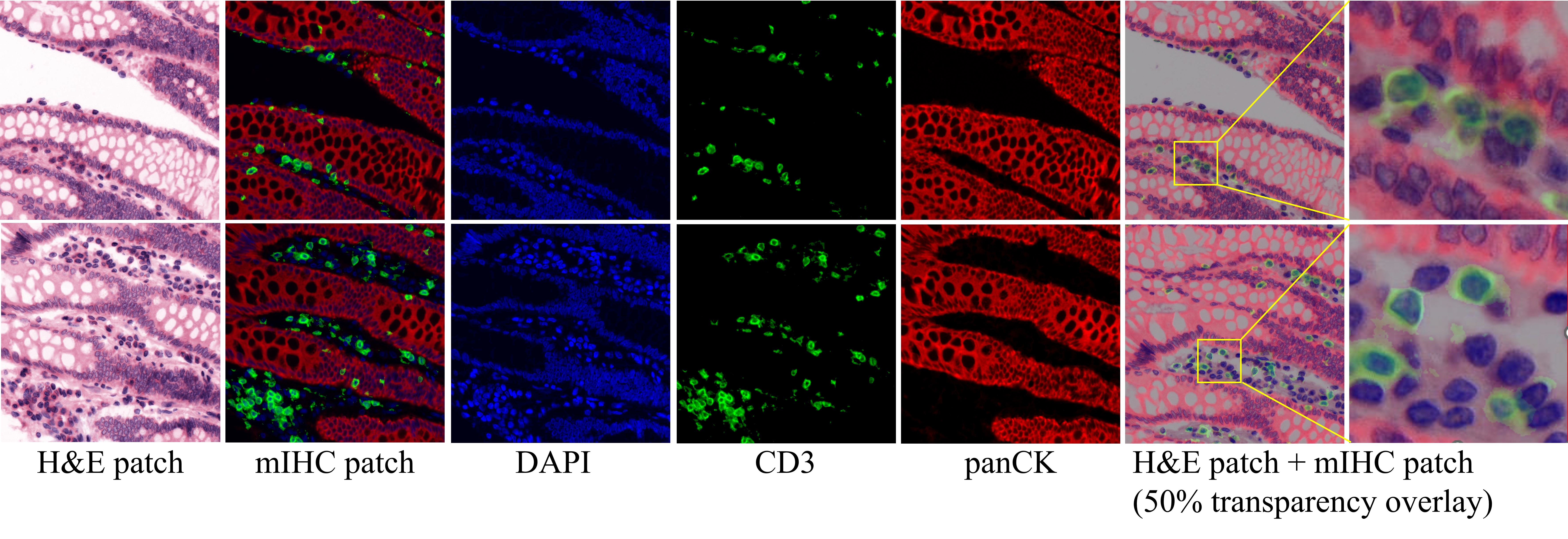} \\
    (a) Visualization of registered HEMIT image pairs and their constituent channels.\\
    \includegraphics[width=1\linewidth]{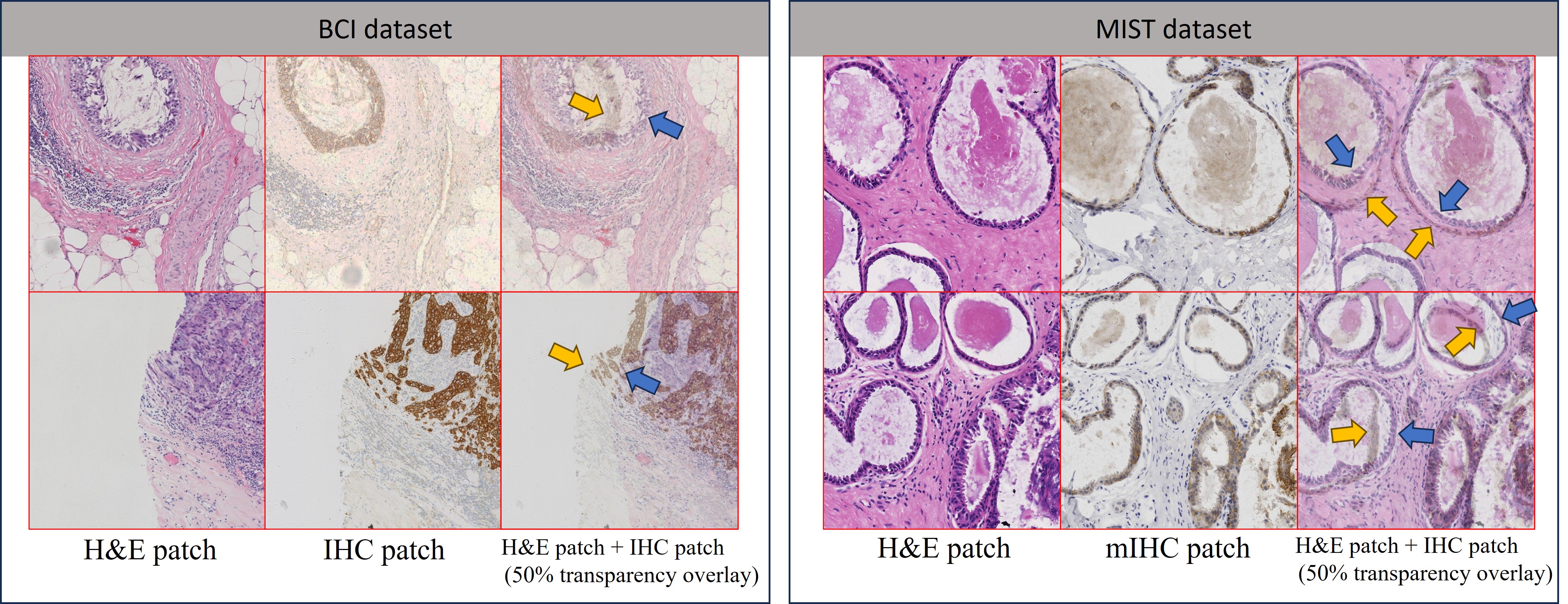} \\
    (b) Visualization of misaligned image pairs from consecutive sectioning datasets.
  \end{tabular}
  \caption{Alignment comparison of HEMIT and other datasets: (a) Visualization of registered image pairs of HEMIT dataset. (b) Visualization of the misaligned image pairs in BCI dataset \cite{liu2022bci} and MIST \cite{li2023adaptive} dataset which used consecutive sectioning. Yellow arrows point to the edges of tissues in HER2 IHC images, and blue arrows point to the edges of tissues in H\&E images to enhance visualization of the misalignment.}
  \label{fig:alignment}
\end{figure}

Upon concluding the registration, a margin of 50 pixels from each edge is trimmed to account for rotational transformations, following methods established in previous work \cite{rivenson2019virtual}. Subsequently, the block pairs are cropped into \(1024 \times 1024\) patches, maintaining a 50\% overlap. Color normalization \cite{macenko2009method} is then applied to all H\&E patches to mitigate stain variations. Visualizations of the final registered image pairs are shown in \cref{fig:alignment}a. This process resulted in 5292 matched image patch pairs. This processed data is distributed into three partitions: \emph{training}, \emph{validation}, and \emph{testing}. The division yields \(3717\) patches designated for training, \(630\) for validation, and \(945\) for the testing phase. Importantly, to prevent data leakage of our evaluation, patches across these subsets are derived from unique patient samples, precluding potential data overlap.

%% file: sec/3_ProposedMethod.tex
\section{Proposed Method}
\label{sec: Proposed Method}
\subsection{Architecture}

\begin{figure}
  \centering
    \includegraphics[width=1\linewidth]{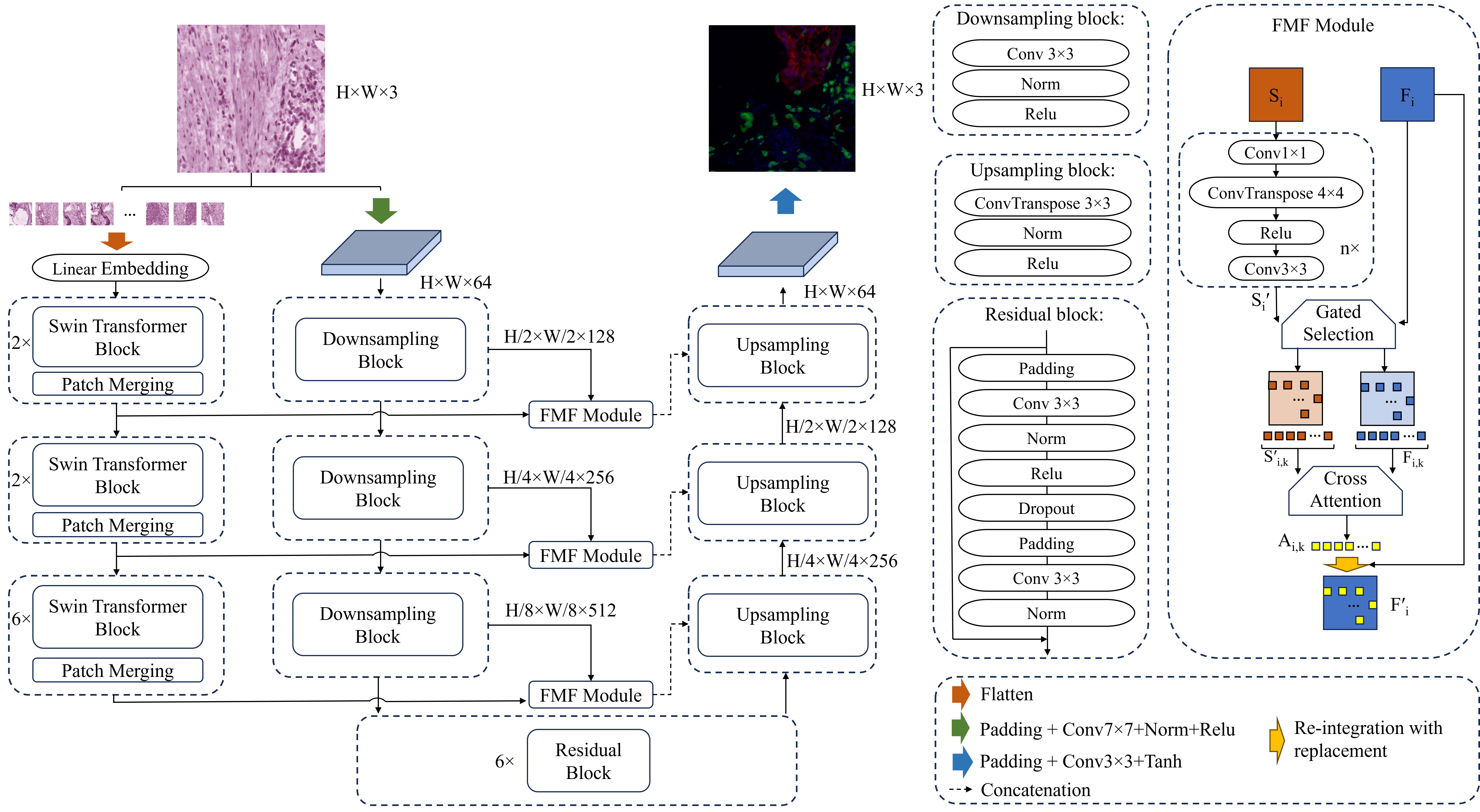}
    \caption{Overall structure of the proposed generator.}
    \label{fig:networkstructure}
\end{figure}

Our dual-branch architecture is based on the original pix2pix \cite{isola2017image} framework, designed with the motivation to fully leverage the detailed information contained within 1024x1024 image patches. These large patches capture comprehensive tissue compartments which provide additional context which has been demonstrated to improve the performance of cellular-level prediction tasks \cite{shephard2021simultaneous}. By incorporating a dual-branch generator, our model can extract multi-scale features from the H\&E inputs. The feature map fusion module then adaptively integrates the information from the 2 branches to boost translation performance. The overall framework is shown in \cref{fig:networkstructure}.

We developed a dual-branch generator architecture. The main branch with residual blocks of the generator is to extract spatial nuances from the input H\&E images, and an auxiliary branch powered by Swin Transformer block \cite{liu2021swin} is incorporated to integrate information across multiple scales. The features extracted by each stage of the Swin Transformer branch are fused with the feature maps of the CNN branch by the Feature Map Fusion (FMF) module. This architecture caters to the multi-scale nature of pathological images. This design captures information from the structure of individual cells to the interaction of cell clusters, vital for interpreting complex tissue environments.

\subsection{Swin Transformer-based Auxiliary Branch}

The auxiliary branch, operating in parallel with the main branch, leverages Swin Transformer modules \cite{liu2021swin} to adeptly capture multi-scale features. Notably, the Swin Transformer introduces both Window-based Multi-head Self Attention (W-MSA) and Shifted Window-based MSA (SW-MSA). Prior to each MSA module and MLP, a Layer Normalization (LN) is applied. $\hat{s}_l$ and $s_l$ denote the outputs of W-MSA and MLP respectively, at the $l^{th}$ layer. These are computed as:
\begin{equation}
\begin{aligned}
    \hat{s}_l &= \text{W-MSA}(\text{LN}(s_{l-1}))+s_{l-1};
    s_l = \text{MLP}(\text{LN}(\hat{s}_l))+\hat{s}_l, \\
    \hat{s}_{l+1} &= \text{SW-MSA}(\text{LN}(\hat{s}_l))+s_l;
    s_{l+1} = \text{MLP}(\text{LN}(\hat{s}_{l+1}))+\hat{s}_{l+1}
\end{aligned}
\end{equation}

The Swin Transformer module is particularly suitable for H\&E images due to its ability to effectively handle different scales of information inherent in these images. H\&E images contain various levels of information, such as cell shape, cell cluster organization, tissue compartments, and cell-stroma interactions, each requiring different levels of resolution for accurate interpretation. The window-based approach of the Swin Transformer allows it to focus on local features within a window, while the shifted window mechanism enables the capture of global context by aggregating information across different windows. This multi-scale feature extraction is crucial for capturing the complex and hierarchical nature of tissue structures in H\&E images, ultimately improving the model's ability to distinguish subtle histological details and interactions.

\subsection{Feature Map Fusion Module}

The FMF module integrates features from the two branches by utilizing Gated Cross Attention, which emphasizes spatially significant regions. For the \(i\)th stage, the Swin Transformer output, \( S_i \), is aligned with the CNN feature map, \( F_i \), in both dimensions and resolution to form \( S'_i \). To identify the most informative regions within \( F_i \), a gating layer is introduced. This gating layer, which can be trained during the training phase, processes \( F_i \) through convolution and sigmoid activation to produce a gating map \( G_i\). The gating map \( G_i\) highlights areas of importance within \(F_i\), enabling the model to focus on spatially significant regions while also managing the computational burden by selectively processing only the most relevant parts of the input.

\begin{equation}
    G_i = \sigma(\text{Conv}(F_i))
\end{equation}

The top \( k \) pixels in \( G_i \) with the highest values represent the most informative regions. Subsets \( F_{i,k} \) and \( S'_{i,k} \) are extracted from \( F_i \) and \( S'_i \) correspondingly for cross attention operation in equation~\ref{eq:attention}. Focusing on these \( k \) points allows the cross-attention to exchange information primarily in regions crucial for the model's performance, thereby improving efficiency. The choice of \( k \) is based on empirical analysis, details in the supplementary materials. The cross attention function is given by:

\begin{equation}
     A_{i,k} = \text{Attention}(F_{i,k}, S'_{i,k}, S'_{i,k}),
     \label{eq:attention}
\end{equation}
where: 
\begin{itemize}
    \item \(F_{i,k}\) serves as the query, representing the top k elements from the CNN feature maps.
    \item \(S'_{i,k}\) serves as the key and value, derived from the corresponding elements in the Swin Transformer output.
\end{itemize}

After the cross-attention operation, the resulting attention matrix \( A_{i,k} \) is re-integrated into the original \( F_i \), replacing original values correspondingly. This re-integration process integrates the spatial context learned from the auxiliary Swin Transformer branch into the primary CNN feature maps, enriching \( F_i \) with complementary tissue spatial information and enhancing the overall feature representation.

\subsection{Loss Functions}

The adversarial loss is adopted from the pix2pix framework:
\begin{equation}
    \mathcal{L}_{cGAN}(G,D) = \mathbb{E}_{x,y}[\log D(x,y)] + \mathbb{E}_{x,z}[\log(1-D(x, G(x,z)))].
\end{equation}
The terms \( x \), \( y \), and \( z \) denote the input H\&E image, the ground truth mIHC image, and random noise, respectively. Additionally, the L1 loss from pix2pix is employed to preserve structural consistency:
\begin{equation}
    \mathcal{L}_1 = \mathbb{E}_{x,y,z}[\|y-G(x,z)\|_1].
\end{equation}
Thus, our overarching objective function becomes:
\begin{equation}
    G^* = \text{arg}\ \underset{G}{\text{min}}\  \underset{D}{\text{max}}\  \mathcal{L}_{cGAN}(G,D) + \lambda \mathcal{L}_1(G),
\end{equation}
where \( \lambda \) balances the contributions of the adversarial loss and the $L_1$ loss.

%% file: sec/4_Experiments.tex
\section{Experiments}

We used the Adam optimizer over 100 epochs, with an initial learning rate of 0.00003 for 50 epochs, and then linearly decayed to zero over the remaining epochs, on an NVIDIA Tesla V100 16GB GPU.

\subsection{Benchmark Results}
Following established benchmarks in image-to-image translation \cite{liu2022bci,burlingame2018shift,christiansen2018silico,liu2020global}, we adopt the Structural Similarity Index Measure (SSIM), Pearson correlation score (R), and Peak Signal to Noise Ratio (PSNR) as evaluation metrics to gauge the quality of the synthesized images.

SSIM assesses the perceived quality by comparing luminance, contrast, and structural information, capturing structural content and visual quality. Pearson correlation score (R) measures the linear correlation between the synthesized and reference images, indicating overall consistency in pixel intensity distribution. PSNR quantifies the ratio between the maximum possible signal power and the power of corrupting noise, reflecting the accuracy and precision of image synthesis. Employing these three metrics provides a comprehensive evaluation, capturing different aspects of image quality for robust performance analysis.

\begin{figure}
  \centering
    \includegraphics[width=1\linewidth]{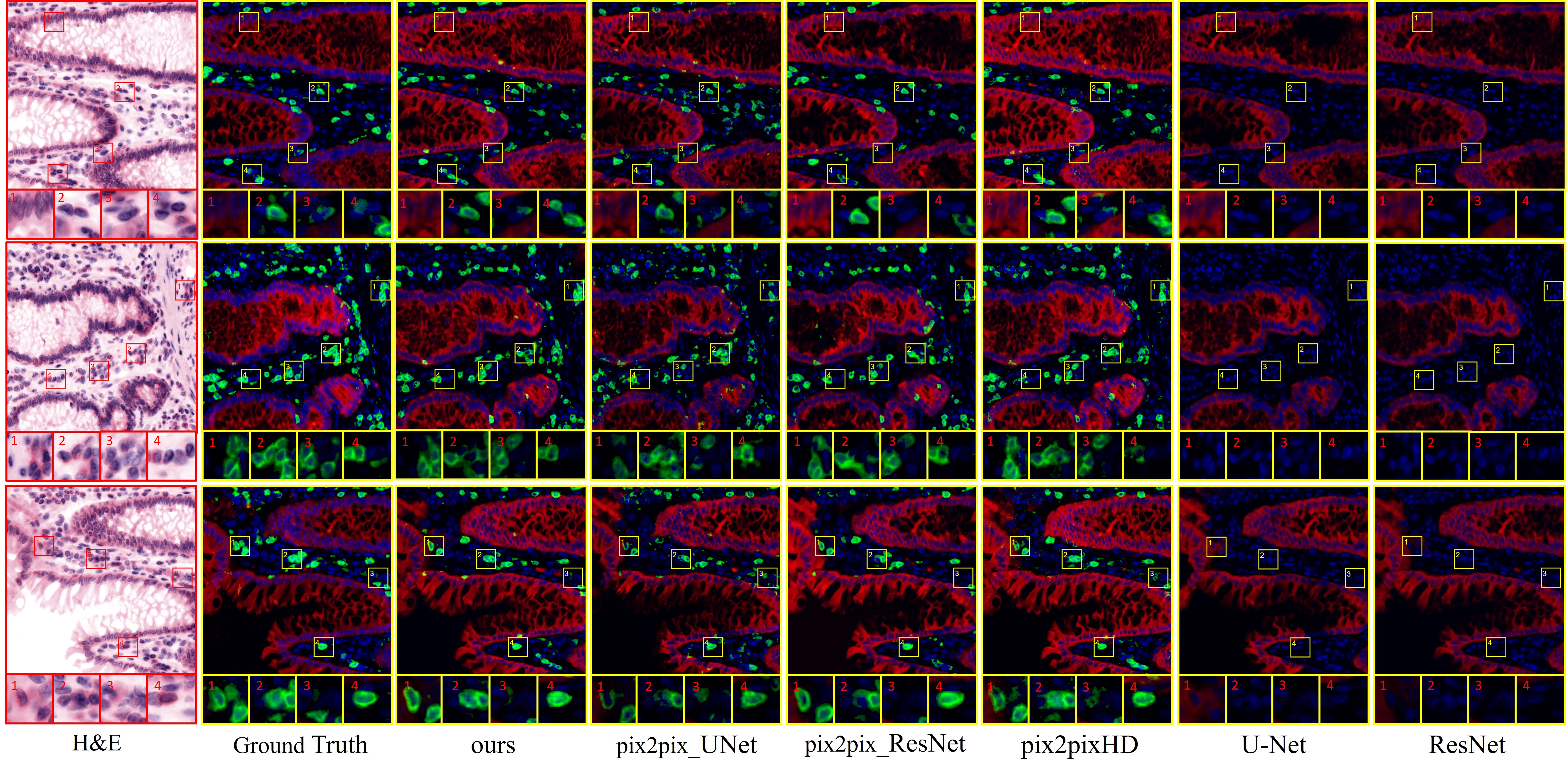}
    \caption{Visualization of different methods on HEMIT dataset. DAPI is shown in blue, panCK in red, and CD3 in green. Below each patch, four zoomed-in regions are displayed to provide a more detailed view.}
    \label{fig:results}
\end{figure}

Our method demonstrated superior performance on the HEMIT dataset, achieving the highest scores in SSIM (0.875), Pearson correlation (0.746), and PSNR (29.886), as shown in \cref{tab:hemit}. This indicates its effectiveness in image translation quality compared to other models. Visualization examples are shown in Figure \ref{fig:results}. The images generated by our methods demonstrate good overall alignment with the ground truth and consistency in marker expression. This indicates that our dual-branch architecture effectively handles stain translation by maintaining structural integrity and marker accuracy across the translated images.

\begin{table}
\caption{Comparison of evaluation metrics across different methods.}\label{tab:hemit}
\centering
{\tiny 
\renewcommand{\arraystretch}{1.3} 
\begin{tabular}{|l|cccc|cccc|cccc|}
\hline
& \multicolumn{4}{c|}{SSIM} & \multicolumn{4}{c|}{R} & \multicolumn{4}{c|}{PSNR (dB)} \\
Methods & DAPI & CD3 & panCK & Average & DAPI & CD3 & panCK & Average & DAPI & CD3 & panCK & Average \\
\hline
U-Net & 0.791 & \textbf{0.907} & 0.901 & 0.866 & 0.659 & 0.005 & 0.949 & 0.538 & 27.929 & 24.733 & 33.773 & 28.695\\
ResNet & 0.790 & \textbf{0.907} & 0.898 & 0.865 & 0.677 & 0.005 & 0.949 & 0.544 & 27.531 & 24.733 & 33.212 & 28.268\\
pix2pix\_UNet & 0.775 & 0.879 & 0.903 & 0.852 & 0.652 & 0.455 & 0.943 & 0.683 & 27.691 & 25.659 & 34.306 & 29.219\\
pix2pix\_ResNet & 0.723 & 0.906 & \textbf{0.914} & 0.848 & 0.668 & 0.553 & 0.946 & 0.723 & 27.152 & 26.225 & 34.359 & 29.189\\
pix2pixHD & 0.786 & 0.900 & 0.888 & 0.858 & \textbf{0.721} & 0.530 & 0.943 & 0.731 & 28.156 & 26.281 & 34.033 & 29.471 \\ 
ours & \textbf{0.815} & 0.898 & 0.913 & \textbf{0.875} & 0.716 & \textbf{0.571} & \textbf{0.951} & \textbf{0.746} & \textbf{28.610} & \textbf{26.349} & \textbf{34.875} & \textbf{29.886}\\
\hline
\end{tabular}
} 
\end{table}

While SSIM is commonly used in previous works regarding stain translation tasks \cite{li2023adaptive,liu2022bci}, it has limitations when used alone. SSIM may offer misleading results, especially in cases where marker expression is highly imbalanced. For instance, as shown in \cref{fig:results}, the UNet and ResNet models completely ignore the CD3 marker in their predictions, focusing only on the dominant markers (DAPI and PanCK). Despite this, these models achieve the highest SSIM scores. This indicates that SSIM alone is insufficient for evaluating the quality of stain translation tasks and highlights the importance of using multiple metrics to obtain a more accurate and reliable assessment.

While U-Net and ResNet scored well in SSIM and PSNR, they fell short in Pearson correlation due to insufficient emphasis on less dominant markers, such as CD3. This highlights the importance of Pearson correlation for evaluating multi-stain translations and demonstrates that GAN-based frameworks significantly improve the realism of image translations, particularly for low-expression markers.

\subsection{Downstream Analysis}

\begin{figure}
  \centering
    \includegraphics[width=1\linewidth]{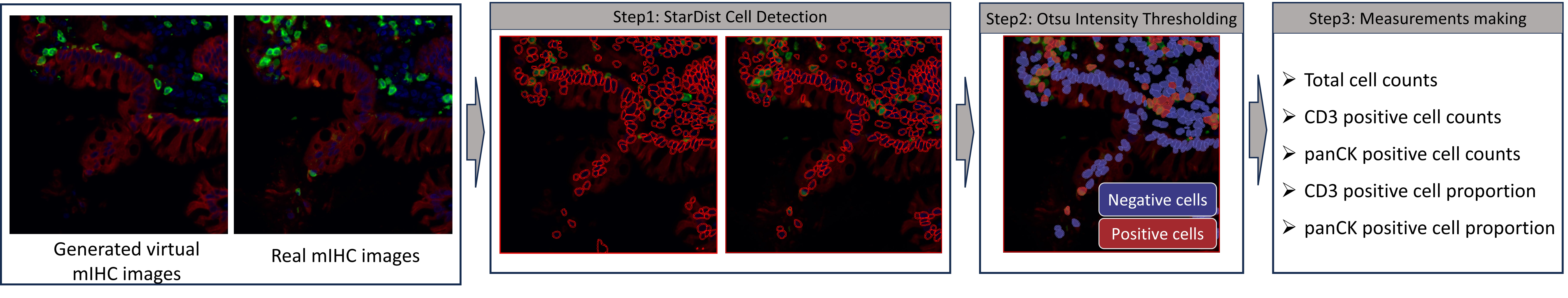}
    \caption{Downstream analysis pipeline: Real and generated virtual mIHC images are analyzed using the Qupath StarDist model for cell detection. Following cell detection, the Otsu thresholding method is applied to identify positive cell groups. Finally, measurements are taken for evaluation purposes.}
    \label{fig:downstream_pipeline}
\end{figure}

To further evaluate the performance of the HEMIT model's generated mIHC images and to give an indication for it's potential utility in clinical studies, we compared the cell overall counts obtained between the generated and real images using an mIHC image analysis pipeline. 
Both real and generated images were analyses using the QuPath application \cite{bankhead2017qupath,loughrey2018validation} along with the nuclear segmentation system Stardist \cite{schmidt2018cell} to identify individual cell nuclei. For each cell, the mean stain expression was extracted for each marker and Otsu segmentation \cite{ostu1979threshold} was applied to identify marker positivity enabling a comparison of the positive cell proportions between real and generated images. A schematic of the downstream analysis pipeline is shown in \cref{fig:downstream_pipeline}.

\begin{figure}
  \centering
    \includegraphics[width=1.0\linewidth]{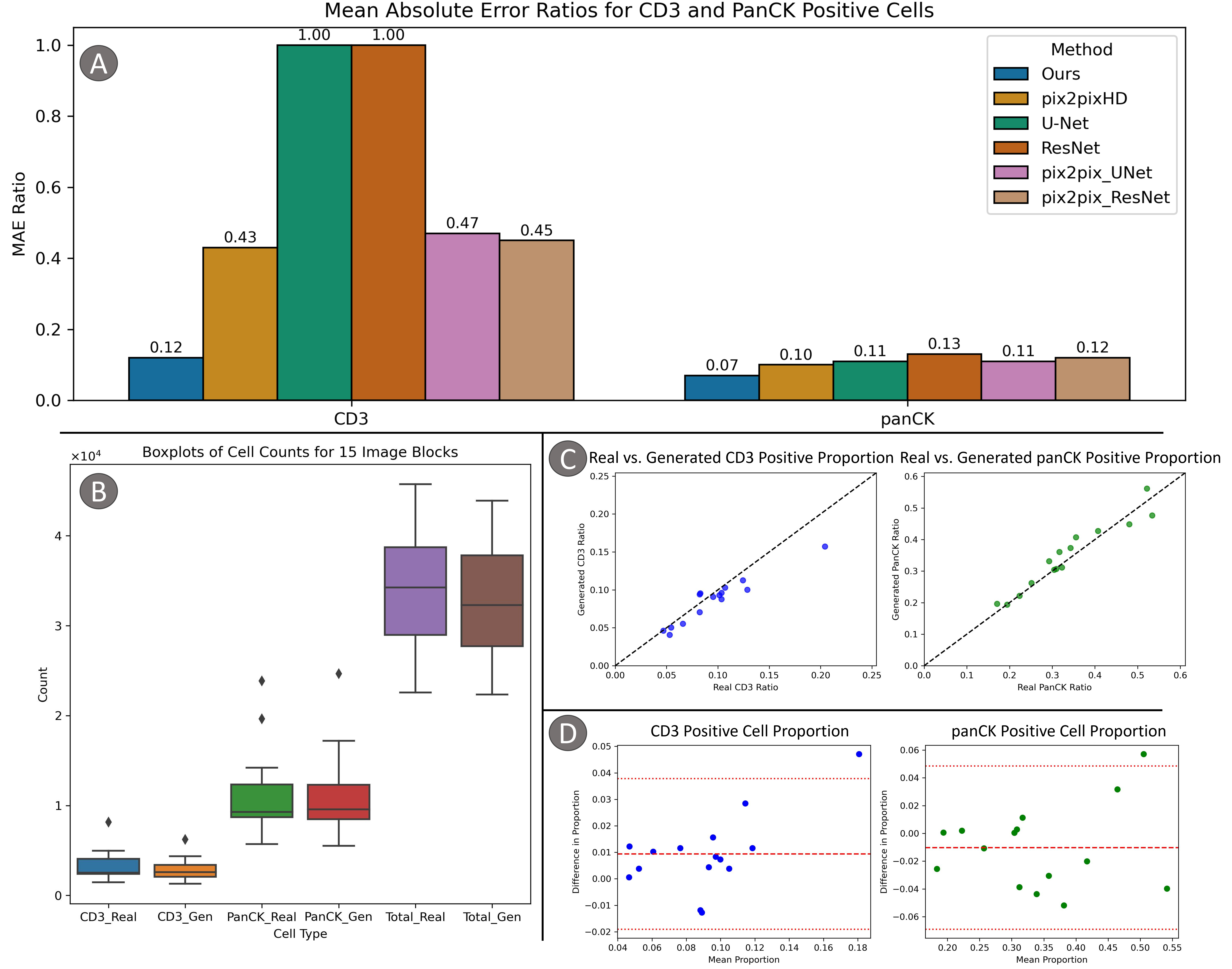}
    \caption{Downstream Analysis Results: (A) Comparative MAE Ratios for CD3 and PanCK; (B) shows boxplots of cell counts for CD3, panCK, and total of the generated images of our method and real images; (C) and (D) provide scatter plots and Bland-Altman plots for CD3 and panCK positive cell proportions of the generated images of our method and the real images.}
    \label{fig:Downstream}
\end{figure}

We calculated the mean absolute error ratios of positive cell proportions for all methods. The mean absolute error ratios are shown in \cref{fig:Downstream}A. Our method shows considerable improvement over the existing state-of-the-art methods, especially on the challenging CD3 marker. Further comparative results of our method are shown in \cref{fig:Downstream}B-D. \cref{fig:Downstream}C and \cref{fig:Downstream}D showcase the close correspondence between real and generated images, evidenced by the tight distribution around the identity line in scatter plots and marginal deviations in Bland-Altman plots.
The mean absolute error ratio is calculated using the formula:
\begin{equation}
\text{Mean Absolute Error Ratio} = \frac{1}{n} \sum_{i=1}^{n} \left| \frac{p_i^{\text{real}} - p_i^{\text{generated}}}{p_i^{\text{real}}} \right|
\end{equation}
where \( p_i^{\text{real}} \) and \( p_i^{\text{generated}} \) represent the positive cell proportions of the real and generated images for the \(i\)-th sample, respectively, and \( n \) is the total number of test samples.

%% file: sec/5_Conclusions.tex
\section{Conclusion}

In this study, we introduced HEMIT, a pioneering dataset tailored for pathology image translation tasks. HEMIT is specifically designed to bridge the gap between H\&E stained sections and their mIHC counterparts, showcasing multiple markers including DAPI, CD3, and panCK. Notably, the mIHC images in our dataset are multi-component, providing a cellular-level registration with the associated H\&E images. This unique alignment presents an invaluable opportunity for supervised stain translation research.

Complementing our dataset contribution, we proposed a dual-branch generator architecture optimized for pathology image analysis. Through extensive experimentation, we contrasted our approach with several SOTA algorithms. The empirical outcomes with downstream analysis underscore the efficacy of our method, establishing a new benchmark for the H\&E to mIHC translation domain. We hypothesise that the significant performance improvement is due to the spatial context provided by the Swin Transformer where the tissue compartment can inform the cell protein expression. This is in line with other work \cite{shephard2021simultaneous} which has demonstrated that tissue context can inform cell classification leading to improved accuracy. In addition, we also discussed the limitation of the commonly used metrics in certain scenarios, and provided a comprehensive evaluation scheme which could offer guidance for future research. 

Moreover, in addition to the metrics commonly used in the computer vision field, we proposed a downstream analysis pipeline from a clinically-focused perspective. The downstream analysis demonstrated that while traditional metrics showed marginal improvements, significant enhancements were observed at the cellular level. This downstream pipeline not only proved the efficacy of our proposed method but also provided a supplementary evaluation approach for stain translation tasks.

This work provides an opportunity for further exploration in the use of predicting biomarkers from H\&E pathology images to accelerate research in biomarker prediction.